\documentclass[]{spie}  

\usepackage{amsmath,amsfonts,amssymb}
\usepackage{graphicx}
\usepackage[colorlinks=true, allcolors=blue]{hyperref}

\title{CONCERTO: Instrument model of Fourier transform spectroscopy, white-noise components}
\author[a,b,c]{Alessandro~Fasano}
\author[d]{Peter~Ade}
\author[e]{Manuel~Aravena}
\author[f]{Emilio~Barria}
\author[a]{Alexandre~Beelen}
\author[f]{Alain~Beno\^it}
\author[a]{Matthieu~B\'ethermin}
\author[g]{Julien~Bounmy}
\author[g]{Olivier~Bourrion}
\author[f]{Guillaume~Bres}
\author[f]{Martino~Calvo}
\author[g]{Andrea~Catalano}
\author[h]{Carlos~De Breuck}
\author[i]{François-Xavier~D\'esert}
\author[a]{Cédric~Dubois}
\author[j]{Carlos~Dur\'an}
\author[a]{Thomas~Fenouillet}
\author[a]{Jose~Garcia}
\author[f]{Gregory~Garde}
\author[f]{Johannes~Goupy}
\author[g]{Christophe~Hoarau}
\author[a]{Wenkai~Hu}
\author[a]{Guilaine~Lagache}
\author[a]{Jean-Charles~Lambert}
\author[f]{Florence~Levy-Bertrand}
\author[a]{Andreas~Lundgren}
\author[g]{Juan-Francisco~Mac\'ias-P\'erez}
\author[g]{Julien~Marpaud}
\author[f]{Alessandro~Monfardini}
\author[d]{Giampaolo~Pisano}
\author[i]{Nicolas~Ponthieu}
\author[a]{Leo~Prieur}
\author[g]{Samuel~Roni}
\author[g]{Sebastien~Roudier}
\author[g]{Damien~Tourres}
\author[a]{Carol~Tucker}
\author[a]{Mathilde~Van~Cuyck}

\affil[a]{Aix Marseille Univ., CNRS, CNES, LAM, Marseille, France}
\affil[b]{Instituto de Astrof\'isica de Canarias, Santa Cruz de Tenerife, E-38205 La Laguna, Spain}
\affil[c]{Departamento de Astrofísica, Universidad de La Laguna (ULL), E-38206 La Laguna, Tenerife, Spain}
\affil[d]{Astronomy Instrumentation Group, University of Cardiff, The Parade, CF24 3AA, United Kindgom}
\affil[e]{N\'ucleo de Astronom\'ia, Facultad de Ingenier\'ia y Ciencias, Universidad Diego Portales, Av.  Ej\'ercito 441, Santiago, Chile}
\affil[f]{Univ. Grenoble Alpes, CNRS, Grenoble INP, Institut N\'eel, 38000 Grenoble, France}
\affil[g]{Univ. Grenoble Alpes, CNRS, LPSC/IN2P3, 38000 Grenoble, France}
\affil[h]{European Southern Observatory, Karl Schwarzschild Straße 2, 85748 Garching, Germany}
\affil[i]{Univ. Grenoble Alpes, CNRS, IPAG, 38000 Grenoble, France}
\affil[j]{European Southern Observatory, Alonso de Cordova 3107, Vitacura, Santiago, Chile}

\authorinfo{Furthert author information: (Send correspondence to Alessandro Fasano)\\Alessandro Fasano: E-mail: alessandro.fasano@iac.es}

\begin{document} 
\maketitle

\begin{abstract}
Modern astrophysics relies on intricate instrument setups to meet the demands of sensitivity, sky coverage, and multi-channel observations. An example is the CONCERTO project, employing advanced technology like kinetic inductance detectors and a Martin-Puplett interferometer. This instrument, installed at the APEX telescope atop the Chajnantor plateau, began commissioning observations in April 2021. Following a successful commissioning phase that concluded in June 2021, CONCERTO was offered to the scientific community for observations, with a final observing run in December 2022. CONCERTO boasts an 18.5\,arcmin field of view and a spectral resolution down to 1.45\,GHz in the 130--310\,GHz electromagnetic band. We developed a comprehensive instrument model of CONCERTO inspired by Fourier transform spectrometry principles to optimize performance and address systematic errors. This model integrates instrument noises, subsystem characteristics, and celestial signals, leveraging both physical data and simulations. Our methodology involves delineating simulation components, executing on-sky simulations, and comparing results with real observations. The resulting instrument model is pivotal, enabling a precise error correction and enhancing the reliability of astrophysical insights obtained from observational data. In this work, we focus on the description of three white-noise noise components included in the instrument model that characterize the white-noise level: the photon, the generation-recombination, and the amplifier noises.
\end{abstract}


\section{INTRODUCTION}
\label{sec:intro}

In the current age of precise astrophysics, our scientific endeavors drive technological innovation, expand the limits of what we can achieve, and propel the development of state-of-the-art technology. As science continually sets higher standards for progress, instruments evolve to meet increasingly stringent specifications, resulting in a parallel increase in complexity.

To navigate this intricate technological landscape, meticulous control is essential for accurately predicting scientific performance and determining the most suitable instrument configurations. Simply considering the subsystems of these instruments in isolation risks drawing flawed conclusions. In this dynamic environment, where complexity permeates every aspect, mathematical modeling of astronomical instruments emerges as an indispensable tool. It serves as a linchpin for technological evolution, aids in the interpretation of data, and enables the characterization, mitigation, and prediction of systematic effects.

In millimetric (mm) astronomy, advancing scientific objectives hinge on two primary imperatives: boosting mapping speed and enhancing spectrometric capability. These objectives are crucial for achieving extensive sky coverage and accessing broadband spectra while effectively separating various emitting components.

The simplest setup for conducting multi-channel observation implies exploiting a dedicated array for a single electromagnetic frequency. However, this approach inevitably inflates the total pixel count or requires a reduction in the field of view. Consequently, such configurations are more suited to studying compact objects than mapping extended fields.

Fourier transform spectroscopy (FTS) emerges as a technique predicated on the principles of Fourier analysis, wherein two beams intersect to generate interference patterns. Unlike methods reliant on dispersive elements to separate light components, FTS produces interferograms, which are subsequently subjected to Fourier transformation, enabling the reconstruction of the observed electromagnetic spectrum. Fourier transform spectroscopy offers versatility and precision, particularly when utilized alongside absolute calibrators, an optimal approach for studying primordial spectral distortions in the cosmic microwave background\cite{FIRAS}. Notably, FTS has garnered attention for its potential deployment in upcoming missions, underscoring its widespread recognition as a pivotal tool in pushing the frontiers of mm astronomy\cite{2017arXiv170602464A}.

Fourier transform spectral mapping fulfills the necessities of area coverage and low-resolution spectroscopy in astronomical observations. Exploiting fast detectors helps overcome the problem of atmospheric instability, envisioning spectral capability while preserving large mapping speed. Coupling FTSs with fast detectors, particularly kinetic inductance detectors (KIDs), offers a promising solution for modern astronomical observation. Kinetic inductance detectors provide state-of-the-art sensitivity and the highest speed in the readout in the mm market.

Cutting-edge incoherent detectors functioning within the mm-wave spectrum typically manifest a predominant noise factor identified as photon noise, stemming from the stochastic behavior of incoming photons. When detectors operate within a range where this criterion is met, they are known as photon-noise limited, achieving optimal noise performance. Subsequently, the next most significant noise source in the white-noise component often emerges as electronic noise introduced by the readout.
In the specific case of KIDs, noise additionally emanates from fluctuations in the superconducting particles' state, termed generation-recombination noise (GR), and fluctuations between the substrate's surface's two energy states, referred to as two-level system noise (TLS). Within the 100--1\,000\,Hz range, the dominant component typically is GR noise, while TLS diminishes with increasing electromagnetic frequency\cite{10.1093/ptep/ptac023}. The achievement of the CONCERTO detectors performance expectation has been substantiated through initial laboratory characterization\cite{2020A&A...641A.179C,2020A&A...642A..60C}, further tests during the commissioning phase\cite{2022JLTP..tmp...51M,2022EPJWC.25700010C}, and astrophysical calibrations (Hu et al., submitted). Leveraging this evidence, assumptions can be made to simplify the instrument model concerning noise contributions, expediting mock data generation that necessitates extensive computational capacity.

The detailed presentation of the instrument model will be reserved for a forthcoming paper. In this paper, we will describe three white-noise components of the model: the photon, the GR, and the amplifier noise.
Section~\ref{sec:photon_noise} describes the photon and GR noise model. We explain the development of the amplifier noise model in Sec.~\ref{sec:amplifiers}. Section~\ref{sec:results} compares on-sky results with the model calculation. Finally, the conclusions are discussed in Sec.~\ref{sec:conclusions}.

\section{PHOTON AND GENERATION-RECOMBINATION NOISE}
\label{sec:photon_noise}
Photon noise arises from the statistical distribution of photons reaching the detector surface, governed by the Bose-Einstein and Poisson distribution. This randomness in arrival times results in fluctuations in the registered power. In the classical formulation of the photon noise equivalent power (NEP$_{\rm{ph}}$), two terms represent the contributions from the Poisson and Bose-Einstein processes\cite{Lamarre:86,Zmuidzinas:03}:

\begin{equation}
	 \rm{ NEP^2_{\rm{ph}}= 
  2h\int_{0}^{\infty }\nu H(\nu) P(\nu)  \,\rm{d}\nu + 2\int_0^\infty \frac{ c^2}{A\Omega\nu^2} \left( H(\nu) P(\nu) \right)^2 \,\rm{d}\nu } \text{ ,}
	\label{eq:NEP}
\end{equation}

\noindent where h is the Planck constant, $\rm{ H(\nu) }$ is the peak-normalized bandpass (dimensionless), P is the optical load, and A$\Omega$ is the system throughput. Firstly, we need to evaluate P to compute the photon noise in the CONCERTO model. We have two different transmission bands in CONCERTO, a low-frequency (LF, 130--270\,GHz) and a high-frequency (HF, 195--310\,GHz), so we have two different optical loads.
The photon noise in CONCERTO comprises two major components: 1) a fixed contribution from the instrument, including eleven aluminum mirrors, polypropylene lenses, and the COLD reference\cite{2020A&A...642A..60C}, and 2) a variable component from the atmosphere, influenced by the opacity and the elevation angle of observation.
The instrument contribution remains fixed because its elements are housed within the temperature-stabilized APEX Cassegrain cabin (C-cabin), maintained at $11.0\pm0.1$\,\textcelsius, with drifts of around 0.5\,\textcelsius, over a 12-hour timescale. The aluminum mirrors exhibit an emissivity of $\sim$1\,\% at mm wavelengths. The lens characteristics are listed in Tab.~\ref{tab:lenses}.

\begin{table}[ht]
        \footnotesize
        \centering
        \caption{Polypropylene lenses characteristics. n is the refracting index. }
        \begin{tabular}{cccc}
        \hline \hline
Lens temperature & Thickness & n \\ \hline
284\,K & 45\,mm & 1.50 \\
4\,K & 15\,mm & 1.52 \\
100\,mK & 45\,mm & 1.52 \\  \hline
        \end{tabular}
        \label{tab:lenses}
\end{table}

\begin{figure} [ht]
   \begin{center}
   \begin{tabular}{cc} 
   \includegraphics[height=8.cm]{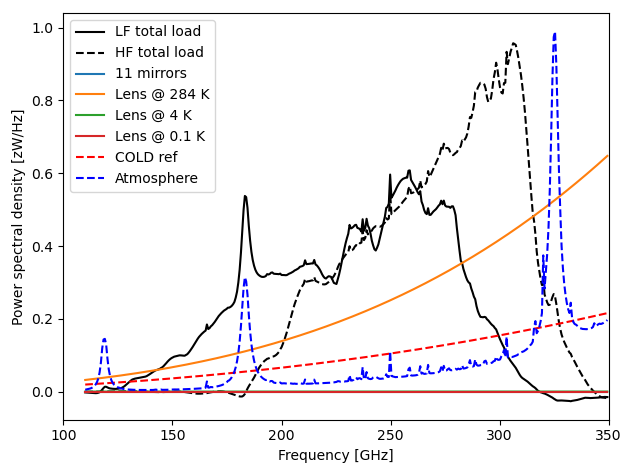}
   \end{tabular}
   \end{center}
   \caption{Optical load from COLD reference (red dashed line), the eleven mirrors (blue line), and the lenses (various colors but the only one that contributes significantly is in orange). A dashed blue line represents the atmospheric emission at 55.7\,deg elevation angle and PWV=0.81\,mm. Water and molecular oxygen emission lines dominate the spectra above the continuum. The total optical loads (the sum of all the components), multiplied by the normalized bandpasses and the optical coupling efficiency, are represented in black continuous and dashed lines, respectively for LF and HF.}
   \label{fig:opt_load}
\end{figure}

For the photon and GR noise calculation, we treat the total optical load as a constant within a single simulated dataset, as typical changes in elevation angle or PWV variations do not drastically alter the overall optical load. The overall load remains relatively stable even under worst-case CONCERTO observed conditions, such as PWV=3\,mm.
The atmospheric emission is calculated using the \texttt{am} atmospheric model at the Chanjantor plateau\cite{paine_scott_2022_6774376}\footnote{\url{https://lweb.cfa.harvard.edu/~spaine/am}}. Figure~\ref{fig:opt_load} illustrates the single contribution of the total optical load, from the COLD reference and the other instrument components, as well as the atmosphere for PWV=0.8\,mm at an elevation angle of 55.7\,deg\footnote{We use these particular values to compare the results with the calibration on-sky at the same atmospheric condition afterward.}.

In addition, to properly calculate the incoming power, we need to include the total optical coupling efficiency ($\eta_{\rm{tot}}$) to correct the optical load, the total coupling factor between the incoming radiation from outside the telescope and the absorbed photon in a single beam (and KID, considering F$\lambda$1), defined as:

\begin{equation}
\rm{ \eta_{tot}= \eta_{mir} \times  \eta_{det}  \times T / 4  \text{ ,}}
\end{equation}

\noindent where $\eta_{\rm{mir}}$ is the fractional reflectivity of the mirrors $\left(0.99 ^{11}\sim0.90 \right)$, $\eta_{\rm{det}}$ is the KID efficiency ($\sim$0.7 from laboratory tests), T is the total transmission from the lenses and filters ($\rm{T_{LF}}=0.79$ and $\rm{T_{HF}}=0.72$ for LF and HF, respectively; Fasano et al., in preparation), and the factor four comes from the two polarizers of the CONCERTO Martin-Puplett FTS\cite{mpi,2020A&A...642A..60C}. For point sources, we also consider the multiplication of $\epsilon_{\rm{M}}$, which is the beam efficiency ($\sim$0.53, from observation on the sky; Hu et al., submitted). The total optical coupling efficiencies for point sources ($\eta_{\rm{opt}}=\eta_{\rm{tot}}\times\epsilon_{\rm{M}}$) are, thus, $\eta\rm{^{LF}_{opt}}=0.05$ and $\eta\rm{^{HF}_{opt}}=0.06$, for LF and HF respectively. These results must be compared to the maximum-achievable total coupling optical efficiency per beam by a Martin-Puplett of $\eta_{\rm{opt}}=0.25$. 

For the total optical load calculation, the lenses are not affected by factor four of the Martin-Puplett (there is just a factor two for one polarizer), and for the COLD reference, and the mirrors, they do not suffer from the mirrors' reflectivity ($\eta_{\rm{mir}}$). By integrating along the electromagnetic frequencies, we compute a total detected optical load of $\sim$59\,pW/beam and $\sim$68\,pW/beam for LF and HF, respectively, with PWV=0.81\,mm at 55.7\,deg of elevation, as reported in Fig.~\ref{fig:opt_load}. 

As in any pair-breaking detector, the fundamental limit to the sensitivity is governed by random generation and recombination of quasiparticle \cite{2011PhRvL.106p7004D,2012JLTP..167..335D}.
Starting from the calculated total optical load, we can compute the NEP contribution from the GR noise (NEP$\rm{_{gr}}$) as\cite{2019NatAs...3..989E}:

\begin{equation}
	\rm{NEP}^2_{\rm{gr}}= 
  \int_{0}^\infty 4 \Delta_{\rm{Al}} H(\nu) P(\nu) / \eta_{\rm{pb}}  \,\rm{d}\nu
	\label{eq:NEP_gr}
\end{equation}

\noindent where $\Delta_{\rm{Al}}$=188\,$\mu$eV is the superconducting gap energy of aluminum, and $\eta_{\rm{pb}}$=0.4 is the pair-breaking efficiency\cite{2014SuScT..27e5012G}.

Once the total NEP from photon and GR noise is determined $\left( \rm{ NEP^{2}_{\rm{tot}}= NEP^2_{\rm{ph}}+NEP^2_{\rm{gr}} } \right)$, we compute its related standard deviation per frequency sample ($\sigma_{\rm{total}}$) as:

\begin{equation}
    \rm{ \sigma_{total} = \sqrt{ \int_{0\,Hz}^{f_s/2} NEP^2_{tot} \, df }/\eta_{opt} =  \sqrt{ f_s/2} \times NEP_{tot} /\eta_{opt} }
    \label{eq:sigma2nep}
\end{equation}

\noindent where $\rm{f_s}$ is the sampling frequency which is 3\,815\,Hz (0.26\,ms) for CONCERTO, with the factor two for Nyquist. We, thus, generate the photon and GR noise timeline by calculating a randomized Gaussian distribution with null mean and a standard deviation of $\sigma_{\rm{total}}$.

\section{AMPLIFIERS NOISE}
\label{sec:amplifiers}

The low-noise cold amplifiers (LNAs) take place on the 4\,K stage of the CONCERTO cryostat and are linked to the KID output via superconducting coaxial cables. This setup aims to minimize the overall noise contribution from the amplification chain.
The noise introduced by the amplifiers possesses distinct characteristics from those described previously; it directly influences the raw KID signal. Therefore, it must be electrically simulated and converted into an equivalent optical signal to assess its impact on observations.
The model that characterizes the electrical properties of a single KID is based on the definition of the scattering parameter $S_{21}$:\cite{Gao}

\begin{equation}
        \rm{ S_{21}(f)= \alpha e^{-2\pi j f t_0} \left[ 1-\frac{\frac{Q_{ \mathrm{res}}}{Q_\mathrm{c}}e^{j\phi_0}}{1+2jQ_{ \mathrm{res}}\left(\frac{f-f_0}{f_0}\right)}\right] } \text{,}
        \label{eq:S21}
\end{equation}

\noindent where $f$ is the bias frequency, $\alpha$ is a complex constant accounting for the gain and phase shift through the system, $e^{-2\pi j f t_0}$ corrects the cable delay of the readout system with a time $\rm{t_0}$, $\rm{Q_c}$ is the coupling quality factor, $\rm{Q_{res}}$ is the total quality factor ($\rm{1/Q_{res} =1/Q_i + 1/Q_c}$ where $\rm{Q_i}$ is the internal quality factor), $\phi_0$ is the resonance phase, and f$\rm_{0}$ is the resonance frequency. The parameters exploited for the simulation are obtained during laboratory tests and the commissioning phase at the telescope; they are reported in Tab.~\ref{tab:kid_ele}.

\begin{table*}[ht]
        \footnotesize
        \centering
        \caption{Kinetic inductance detector characteristics of CONCERTO. The responsivity ($\mathfrak{R}$) distinguishes the two arrays at low frequency (LF) and high frequency (HF). }
        \begin{tabular}{cccccccc}
        \hline \hline
$\alpha\rm_0$ & t$\rm_0$ & $\phi\rm{_{0}}$ & f$\rm_{0}$ & Q$\rm_{i}$ {\cite{2022JLTP..tmp...51M}}  & Q$\rm_{c}$ {\cite{2020A&A...642A..60C}} & C & $\mathfrak{R}$ LF/HF \\ \hline
1 & 0\,s & 0 rad & 2\,GHz & 17\,000 & 23\,000 & 150\,kHz/rad & 25.6/19.5\,Hz/Jy \\ \hline
        \end{tabular}
        \label{tab:kid_ele}
\end{table*}

The RMS voltage noise (Johnson-Nyquist) of the amplifier is given by:

\begin{equation}
   \rm{ V_{RMS}^{tot} = \sqrt{ 4 f_s k_B Z T_a} }   
\end{equation}

\noindent where $\rm{k_B}$ is the Boltzmann constant, Z is the nominal impedance (50\,$\Omega$) and T$_{\rm{a}}$ is the amplifier temperature (4\,K). By assuming an equal distribution of the voltage noise on the In-phase (I) and Quadrature (Q) signals (denoted in the rectangular form of the S$_{21}$ signal), we can express the two root mean square (RMS) voltages as:

\begin{equation}
    \rm{ V_{RMS}^I=V_{RMS}^Q=  V_{RMS}^{tot}/\sqrt{2} }
\end{equation}

The CONCERTO amplifiers operate with an input power (P$\rm_{in}$) of -70 dBm. We convert P$\rm_{in}$ to volts using V$\rm_{in}=\sqrt{ P\rm_{in} Z}$. Thus, we can compute the S$_{21}$ signal in voltage by multiplying it by $V\rm_{in}$.
The RMS in terms of I and Q (I$\rm_{RMS}$, Q$\rm_{RMS}$) is then calculated by adding ($\rm{ V_{RMS}^I,V_{RMS}^Q) }$ to the S$_{21}$ at the resonance frequency, S$\rm{_{21}(f_0)=(I_0,Q_0) }$, and then multiplied by the CONCERTO amplifiers' gain factor $\mathfrak{A}$ = 30\,dB:

\begin{equation}
    \rm{ (I\rm_{RMS}, Q\rm_{RMS}) = \mathfrak{A} \times \left[ \left(V_{\rm{RMS}}^I,V_{\rm{RMS}}^Q \right) + \left( I_0,Q_0 \right) \right] }
\end{equation}

In CONCERTO, the I and Q signals are recorded as raw signals. The phase signal ($\phi$) is computed as the arctangent of (I, Q) and then converted into the physical signal: the resonance frequency, F$\rm_{RMS}$ \cite{fasano_aa,2022JInst..17P8037B}.
Kinetic inductance detectors maintain a linear response over a considerable background variation, demonstrated to a temperature signal gradient up to 250\,K\cite{2014JLTP..176..787M}.
Therefore, we can reasonably consider a constant conversion factor (C \cite{fasano_aa}) between the computed phase signal and the resonance frequency. Consequently, a continuous photometric responsivity ($\mathfrak{R}$) converts the resonance frequency signal to flux density.
We obtained C from laboratory measurements and $\mathfrak{R}$ during the photometric assessment on sky (Hu et al., submitted; reported in Tab.~\ref{tab:kid_ele}).
The standard deviation per frequency sample from the amplifiers ($\sigma_{\rm{amp}}$) is, thus:

\begin{equation}
    \rm{\sigma}_{\rm{amp}} = F\rm_{RMS} / \mathfrak{R}
      = \phi\rm_{RMS} \times C / \mathfrak{R}
      = \left[ \arctan( I\rm_{RMS}, Q\rm_{RMS} ) - \arctan( I_0,Q_0 ) \right] \times C / \mathfrak{R}  \text{.} 
      \label{eq:sigma_amp}
\end{equation}

\section{RESULTS}
\label{sec:results}

We present and compare the sensitivity calculation from the instrument model and the results on sky.
Table~\ref{tab:nep} reports all the results for the NEP with the various components.

\begin{table}[ht]
        \footnotesize
        \centering
        \caption{Theoretical noise equivalent power per KID and noise component. The atmosphere is simulated with PWV=0.81\,mm and elevation 55.7\,deg. Photon (ph), generation-recombination (gr), and amplifier (amp) noises.}
        \begin{tabular}{ccccc}
        \hline \hline
 & NEP$\rm{_{ph}}$ $\left[\rm{aW}/\sqrt{\rm{Hz}}\right]$ & NEP$\rm{_{gr}}$ $\left[\rm{aW}/\sqrt{\rm{Hz}}\right]$ & NEP$\rm{_{amp}}$ $\left[\rm{aW}/\sqrt{\rm{Hz}}\right]$ & NEP$\rm{_{total}}$ $\left[\rm{aW}/\sqrt{\rm{Hz}}\right]$ \\ \hline
LF & 304 & 167 & 216 & 409 \\
HF & 329 & 180 & 180 & 416 \\ \hline
        \end{tabular}
        \label{tab:nep}
\end{table}

The primary contributors to the overall white noise are the combined photon and GR noise, with the amplifier also playing a significant role. The single KID is predicted to have, thus, $\rm{NEP}\sim4\times 10^{-16}$\,W/$\sqrt{\rm{Hz}}$ while for the single LF and HF array (2\,152 KIDs per array with $>$72\% validated on sky; Hu et al., submitted) is translated into a $\rm{NEP}\sim1\times10^{-17}$\,W/$\sqrt{\rm{Hz}}$ in agreement with the state-of-the-art performance for wide-band KIDs observing from on ground.

Finally, we can convert the NEP in noise equivalent flux density (NEFD$\rm{_{tot}}$) per beam, which is a more convenient quantity in astronomy, as:

\begin{equation}
    \rm{NEFD_{tot} = \frac{NEP_{total}}{A\times \Delta\nu_{\rm{H}} \times \eta_{opt}} }
\end{equation}

\noindent where A is the APEX illuminated area $\left( 95\,\rm{m}^2 \right)$ and $\Delta\nu_{\rm{H}}$ is the weighted bandwidth as:

\begin{equation}
    \rm{ \Delta\nu_{\rm{H}} = \int_{\Delta \nu} H(\nu) \,d\nu \text{ , } }
\end{equation}

\noindent where $\Delta\nu$ is the electromagnetic frequency range defined within the full width at half maximum of the dimensionless $\rm{ H(\nu) }$ (approximated to a Gaussian distribution). $\Delta\nu_{\rm{H}}$ is 126\,GHz and 93\,GHz for LF and HF, respectively. Table~\ref{tab:nefd} reports the results for the NEFD calculation from the instrument model.

\begin{table}[ht]
        \footnotesize
        \centering
        \caption{Theoretical noise equivalent flux density and standard deviation (per sample at f$_{\rm{s}}$=3\,815\,Hz) per component contribution. Photon (ph), generation-recombination (gr), and amplifier (amp) noises.}
        \begin{tabular}{ccccc}
\hline \hline
 & NEFD$\rm{_{ph}}$ $\left[\rm{mJy}\sqrt{\rm{s}}/\rm{beam}\right]$ & NEFD$\rm{_{gr}}$ $\left[\rm{mJy}\sqrt{\rm{s}}/\rm{beam}\right]$ & NEFD$\rm{_{amp}}$ $\left[\rm{mJy}\sqrt{\rm{s}}/\rm{beam}\right]$ & NEFD$\rm{_{tot}}$ $\left[\rm{mJy}\sqrt{\rm{s}}/\rm{beam}\right]$ \\ \hline
LF & 54.7 & 30.1 & 38.9 & 73.6 \\
HF & 88.1 & 48.2 & 48.1 & 111.3 \\ \hline
 & $\sigma\rm{_{ph}} \, \left[Jy/\rm{beam}\right]$ & $\sigma\rm{_{gr}} \, \left[Jy/\rm{beam}\right]$ & $\sigma\rm{_{amp}} \, \left[Jy/\rm{beam}\right]$ & $\sigma_{\rm{tot}}$ $\left[\rm{Jy}/\rm{beam}\right]$ \\ \hline
LF & 2.4 & 1.3 & 1.7 & 3.2 \\
HF & 3.8 & 2.1 & 2.1 & 4.9 \\ \hline
        \end{tabular}
        \label{tab:nefd}
\end{table}

We can compare the NEFD calculations in Tab.~\ref{tab:nefd} with the continuum on-sky measurements. From on-sky calibration of COSMOS observations, we obtained 95$\pm$1\,mJy$\rm{\sqrt{s}}$/beam and 115$\pm$2\,mJy$\rm{\sqrt{s}}$/beam for LF and HF, respectively, and for PWV=0.81$\pm$0.63\,mm and an elevation of 55.7$\pm$10.8\,deg. For AS2UDS observations, we obtained 111$\pm$7\,mJy$\rm{\sqrt{s}}$/beam and 119$\pm$5\,mJy$\rm{\sqrt{s}}$/beam for LF and HF, respectively, and for PWV=0.75$\pm$0.10\,mm and an elevation of 62.5$\pm$12.4\,deg (Hu et al., submitted). These results should be compared to the NEFD$_{\rm{tot}}$ results in Tab.~\ref{tab:nefd}, which are $\sim$74\,mJy$\rm{\sqrt{s}}$/beam and $\sim$111\,mJy$\rm{\sqrt{s}}$/beam for LF and HF, respectively.

The on-sky NEFD appears to be higher and its variation within similar atmospheric conditions is compatible but has significant uncertainty. The discrepancies might arise from residual 1/f noise subtraction (not accounted for in the calculations of the theoretical NEFD) and signal leakage during noise filtering. In addition, the on-sky NEFD is calculated with the presence of the interferograms in the timeline as well as the known beamsplitter vibration systematics that introduces a background variation in the timeline \cite{2022SPIE12190E..0QF}. Additionally, other sources of mismatch include various assumptions about $\eta_{\rm{det}}$, $\eta_{\rm{mir}}$, $\eta_{\rm{pb}}$, T, and $\Delta \nu _{\rm{H}}$, as well as the amplifier noise characteristics of a CONCERTO KID (Tab.~\ref{tab:kid_ele}), which do not incorporate, for example, the variation of Q$_{\rm{i}}$ with the optical load.
Finally, in the NEFD computation, we do not integrate TLS noise, phonon noise, or account for signal dispersion in the feedline, as well as the optical load coming from the filter and, especially, the two FTS polarizers and the beamsplitter which would increase the total NEFD.

\section{CONCLUSIONS}
\label{sec:conclusions}

We developed an instrument model capable of producing mock data of a high-mapping speed FTS. We adapted this numerical method to the CONCERTO instrument and exploited it to reproduce its behavior regarding instrument characteristics and sky observations. In this work, we described three white-noise components of this model, that produce the randomized timeline of the photon, GR, and amplifier noises and we calculate their NEFD. We compared the results with the on-sky performance.

The on-sky NEFD values are higher than the calculated values.
The difference between the calculated NEFD and the one calibrated on-sky with the maps is $\sim$22\,\% and $\sim$3\,\% for LF and HF, respectively.
Discrepancies may be due to assumptions about the physics characteristics of the instrument that lead to an underestimation of the total optical load. In particular, for LF, the optical load seems to be underestimated.

The instrument model provides a robust framework for simulating CONCERTO's performance and can be refined with further data characterization and insights. Finally, this study will be useful to further instrument optimization for future experiments.

\acknowledgments 
The KID arrays described in this paper have been produced at the PTA Grenoble microfabrication facility.
This work has been partially supported by the LabEx FOCUS ANR-11-LABX-0013, the European Research Council (ERC) under the European Union's Horizon 2020 research and innovation program (project CONCERTO, grant agreement No 788212), and the Excellence Initiative of Aix-Marseille University-A*Midex, a French ``Investissements d'Avenir'' program.
This research made use of {\tt Astropy} (\url{http://www.astropy.org}), a community-developed core Python package for Astronomy\cite{astropy:2013, astropy:2018}. We also use  {\tt Matplotlib} (\url{https://matplotlib.org}\cite{Hunter:2007}), {\tt NumPy} (\url{https://numpy.org}\cite{harris2020array}) and {\tt SciPy} (\url{http://www.scipy.org}, \cite{2020SciPy-NMeth}).

\bibliography{report} 
\bibliographystyle{spiebib} 

\end{document}